 \pdfoutput=1 

\documentclass[10pt,twocolumn]{article}
\usepackage[latin5]{inputenc}
\usepackage{graphicx}
\usepackage{latexsym}
\usepackage{txfonts}
\usepackage{color}
\usepackage{times}
\usepackage{array}
\usepackage{wasysym}
\usepackage{xspace}
\usepackage{setspace}
\usepackage{ulem}
\usepackage{longtable}

\def\Msun{M_{\odot}}

\def\MJ{M_\mathrm{J}}
\def\Mearth{M_{\oplus}}

\def\epsdq{{\varepsilon\,(\log d, \log q)}}
\def\epsam{{\varepsilon\,(\log\apla, \log\mpla)}}

\def\mpla{M}
\def\apla{a}

\def\dd{\mathrm{d}}

\begin{document}

 \onecolumn

 \noindent\textbf{\Huge One or more bound planets per Milky Way star
   from microlensing observations}

 \vspace{0.8cm} 
 \onehalfspacing 
 
 {\noindent\large A.~Cassan$^{1,2,3}$, D.~Kubas$^{1,2,4}$,
   J.-P.~Beaulieu$^{1,2}$, M.~Dominik$^{1,5}$, K.~Horne$^{1,5}$,
   J.~Greenhill$^{1,6}$, J.~Wambsganss$^{1,3}$, J.~Menzies$^{1,7}$,
   A.~Williams$^{1,8}$, U.G.~Jørgensen$^{1,9}$, A.~Udalski$^{10,11}$,
   D.P.~Bennett$^{1,12}$, M.D.~Albrow$^{1,13}$, V.~Batista$^{1,2}$,
   S.~Brillant$^{1,4}$, J.A.R.~Caldwell$^{1,14}$, A.~Cole$^{1,6}$,
   Ch.~Coutures$^{1,2}$, K.H.~Cook$^{1,15}$, S.~Dieters$^{1,6}$,
   D.~Dominis Prester$^{1,16}$, J.~Donatowicz$^{1,17}$,
   P.~Fouqu{\'e}$^{1,18}$, K.~Hill$^{1,6}$, N.~Kains$^{1,19}$,
   S.~Kane$^{1,20}$, J.-B.~Marquette$^{1,2}$, R.~Martin$^{1,8}$,
   K.R.~Pollard$^{1,13}$, K.C.~Sahu$^{1,14}$, C.~Vinter$^{1,9}$,
   D.~Warren$^{1,6}$, B.~Watson$^{1,6}$, M.~Zub$^{1,3}$,
   T.~Sumi$^{21,22}$, M.K.~Szyma{\'n}ski$^{10,11}$,
   M.~Kubiak$^{10,11}$, R.~Poleski$^{10,11}$, I.~Soszynski$^{10,11}$,
   K.~Ulaczyk$^{10,11}$, G.~Pietrzy{\'n}ski$^{10,11,23}$ \&
   {\L}.~Wyrzykowski$^{10,11,24}$}

 \singlespacing
 
 \noindent \textbf{Received 19 January; accepted 28 October 2011.}

% \begin{center} {\vspace{0.2cm}
%    \color{red} \Huge\bf EMBARGOED UNTIL \\ \vspace{0.3cm}
%    11 January 2012, 19:00 CET 
% \vspace{0.2cm}}
% \end{center} 
 \vspace{1.cm}

 \begin{enumerate}\itemsep-2pt
 
 \item Probing Lensing Anomalies Network (PLANET) Collaboration.
 \item Institut d'Astrophysique de Paris, Universit{\'e} Pierre \&
   Marie Curie, UMR7095 UPMC--CNRS 98 bis boulevard Arago, 75014
   Paris, France.
 \item Astronomischen Rechen-Instituts (ARI), Zentrum für Astronomie,
   Heidelberg University, Mönchhofstrasse. 12-14, 69120 Heidelberg,
   Germany.
 \item European Southern Observatory, Alonso de Cordoba 3107,
   Vitacura, Casilla 19001, Santiago, Chile.
 \item Scottish Universities Physics Alliance (SUPA), University of St
   Andrews, School of Physics \& Astronomy, North Haugh, St Andrews,
   KY16 9SS, UK.
 \item University of Tasmania, School of Maths and Physics, Private
   bag 37, GPO Hobart, Tasmania 7001, Australia.
 \item South African Astronomical Observatory, PO Box 9 Observatory
   7935, South Africa.
 \item Perth Observatory, Walnut Road, Bickley, Perth 6076, Australia.
 \item Niels Bohr Institute and Centre for Star and Planet Formation,
   Juliane Mariesvej 30, 2100 Copenhagen, Denmark.
 \item Optical Gravitational Lensing Experiment (OGLE) Collaboration.
 \item Warsaw University Observatory. Al. Ujazdowskie 4, 00-478
   Warszawa, Poland.
 \item University of Notre Dame, Physics Department, 225 Nieuwland
   Science Hall, Notre Dame, Indiana 46530, USA.
 \item University of Canterbury, Department of Physics \& Astronomy,
   Private Bag 4800, Christchurch 8140, New Zealand.
 \item Space Telescope Science Institute, 3700 San Martin Drive,
   Baltimore, Maryland 21218, USA.
 \item Institute of Geophysics and Planetary Physics, Lawrence
   Livermore National Laboratory, PO Box 808, California 94550, USA.
 \item Department of Physics, University of Rijeka, Omladinska 14,
   51000 Rijeka, Croatia.
 \item Technical University of Vienna, Department of Computing,
   Wiedner Hauptstrasse 10, 1040 Vienna, Austria.
 \item Laboratoire d'Astrophysique de Toulouse (LATT), Université de
   Toulouse, CNRS, 31400 Toulouse, France.
 \item European Southern Observatory Headquarters,
   Karl-Schwarzschild-Strasse. 2, 85748 Garching, Germany.
 \item NASA Exoplanet Science Institute, Caltech, MS 100-22, 770 South
   Wilson Avenue, Pasadena, California 91125, USA.
 \item Microlensing Observations in Astrophysics (MOA)
   Collaboration. 
 \item Department of Earth and Space Science, Osaka University, Osaka
   560-0043, Japan.
 \item Universidad de Concepcion, Departamento de Fisica, Casilla
   160-C, Concepción, Chile.
 \item Institute of Astronomy, University of Cambridge, Madingley
   Road, Cambridge CB3 0HA, UK.
 \end{enumerate}

 % -----------------------------
 % Letter - First paragraph
 % -----------------------------

 \twocolumn

 \begin{bf}	
   \noindent
   Most known extrasolar planets (exoplanets) have been discovered
   using the radial velocity$^{\bf 1,2}$ or transit$^{\bf 3}$
   methods. Both are biased towards planets that are relatively close
   to their parent stars, and studies find that around 17--30\% (refs
   4, 5) of solar-like stars host a planet. Gravitational
   microlensing$^{\bf 6\rm{\bf -}\bf 9}$, on the other hand, probes
   planets that are further away from their stars. Recently, a
   population of planets that are unbound or very far from their stars
   was discovered by microlensing$^{\bf 10}$. These planets are at
   least as numerous as the stars in the Milky Way$^{\bf 10}$. Here we
   report a statistical analysis of microlensing data (gathered in
   2002--07) that reveals the fraction of bound planets 0.5--10 AU
   (Sun--Earth distance) from their stars.  We find that 17$_{\bf
     -9}^{\bf +6}$\% of stars host Jupiter-mass planets (0.3--10 $\MJ$,
   where $\MJ {\bf = 318}$ $\Mearth$ and $\Mearth$ is Earth's mass).
   Cool Neptunes (10--30 $\Mearth$) and super-Earths (5--10 $\Mearth$)
   are even more common: their respective abundances per star are
   52$_{\bf -29}^{\bf +22}$\% and 62$_{\bf -37}^{\bf +35}$\%.  We
   conclude that stars are orbited by planets as a rule, rather than
   the exception.
 \end{bf}
 
 ~ % insert newline
 
 % -----------------------------
 % Letter - Corpus
 % -----------------------------
 
 Gravitational microlensing is very rare: fewer than one star per
 million undergoes a microlensing effect at any time. Until now, the
 planet-search strategy$^7$ has been mainly split into two
 levels. First, wide-field survey campaigns such as the Optical
 Gravitational Lensing Experiment (OGLE; ref. 11) and Microlensing
 Observations in Astrophysics (MOA; ref. 12) cover millions of stars
 every clear night to identify and alert the community to newly
 discovered stellar microlensing events as early as possible. Then,
 follow-up collaborations such as the Probing Lensing Anomalies Network
 (PLANET; ref. 13) and the Microlensing Follow-Up Network (mFUN; refs
 14, 15) monitor selected candidates at a very high rate to search for
 very short-lived light curve anomalies, using global networks of
 telescopes.
 
 To ease the detection-efficiency calculation, the observing strategy
 should remain homogeneous for the time span considered in the
 analysis. As detailed in the Supplementary Information, this
 condition is fulfilled for microlensing events identified by OGLE and
 followed up by PLANET in the six-year time span 2002--07. Although a
 number of microlensing planets were detected by the various
 collaborations between 2002 and 2007 (Fig.~1), only a subset of them
 are consistent with the PLANET 2002--07 strategy. This leaves us with
 three compatible detections: OGLE 2005-BLG-071Lb (refs 16, 17) a
 Jupiter-like planet of mass $M \simeq 3.8\ \MJ$ and semi-major axis
 $a \simeq 3.6$~AU; OGLE 2007-BLG-349Lb (ref. 18), a Neptune-like
 planet ($M \simeq 0.2\ \MJ$, $a \simeq 3$~AU); and the super-Earth
 planet OGLE 2005-BLG-390Lb (refs 19, 20; $M \simeq 5.5\ \Mearth$, $a
 \simeq 2.6$~AU).
 
 % ----------------------------------------------------------
 % FIGURE 1 
 % ----------------------------------------------------------
 \begin{figure}[!h]
   \includegraphics[width=\columnwidth]{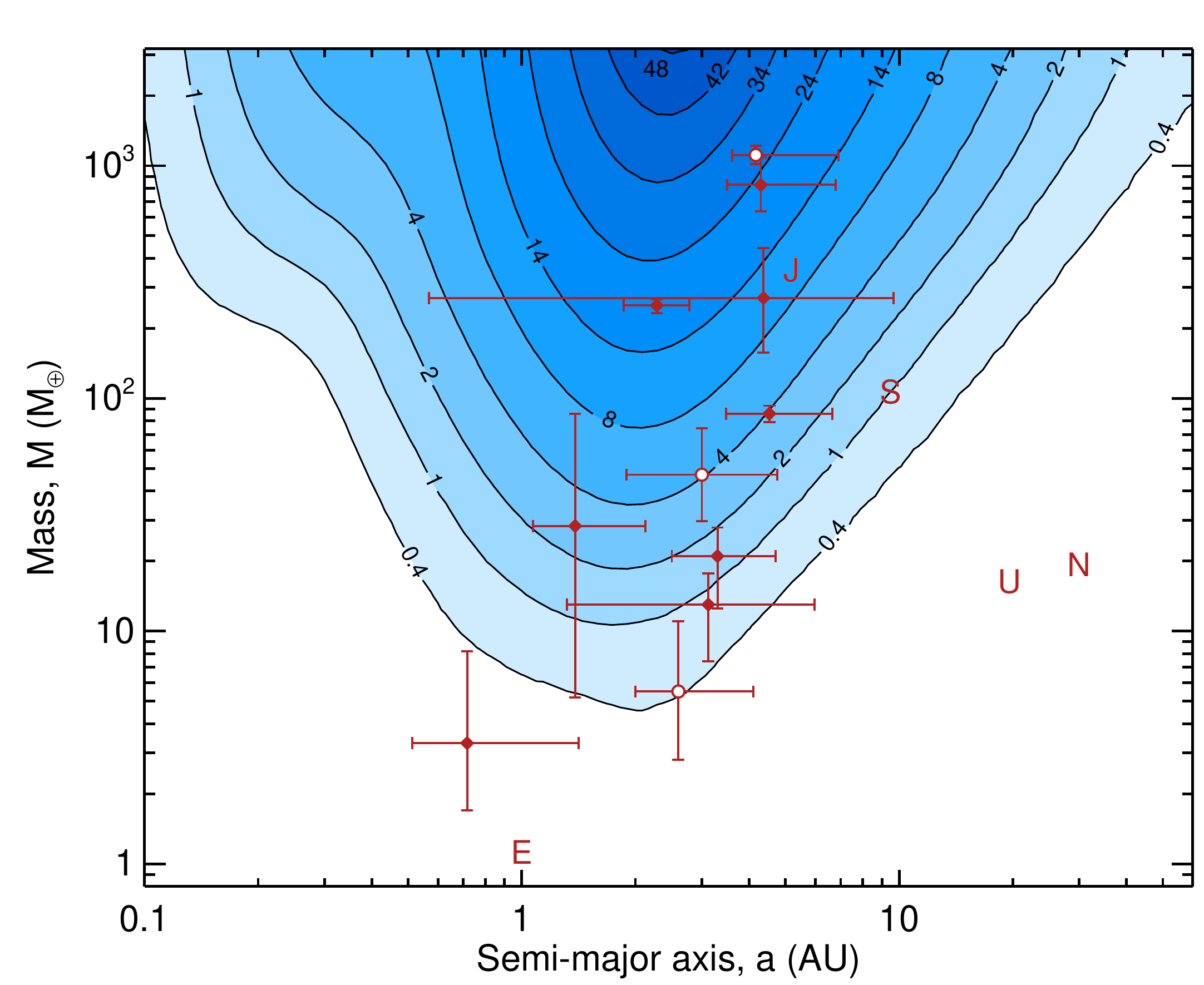}
   \caption{\textbf{Survey-sensitivity diagram.}  Blue contours,
     expected number of detections from our survey if all lens stars
     have exactly one planet with orbit size $a$ and mass $M$. Red
     points, all microlensing planet detections in the time span
     2002--07, with error bars (s.d.) reported from the
     literature. White points, planets consistent with PLANET
     observing strategy. Red letters, planets of our Solar System,
     marked for comparison: E, Earth; J, Jupiter; S, Saturn; U,
     Uranus; N, Neptune. This diagram shows that the sensitivity of
     our survey extends roughly from 0.5~AU to 10~AU for planetary
     orbits, and from 5~$\Mearth$ to 10~$\MJ$. The majority of all
     detected planets have masses below that of Saturn, although the
     sensitivity of the survey is much lower for such planets than for
     more massive, Jupiter-like planets. Low-mass planets are thus
     found to be much more common than giant planets.}
   \label{fig:cassan_fig1.pdf}
 \end{figure}
 % ----------------------------------------------------------

 To compute the detection efficiency for the 2002--07 PLANET seasons,
 we selected a catalogue of unperturbed (that is, single-lens-like)
 microlensing events using a standard procedure$^{21}$, as explained
 in the Supplementary Information. For each light curve, we defined
 the planet-detection efficiency $\epsdq$ as the probability that a
 detectable planet signal would arise if the lens star had one
 companion planet, with mass ratio $q$ and projected orbital
 separation $d$ (in Einstein-ring radius units; ref. 22). The
 efficiency was then transformed$^{23}$ to $\epsam$. The survey
 sensitivity $S(\log a,\log M)$ was obtained by summing the detection
 efficiencies over all individual microlensing events. It provided the
 number of planets that our survey would expect to detect if all lens
 stars had exactly one planet of mass $M$ and semi-major axis $a$.

 We used 2004 as a representative season from the PLANET survey. Among
 the 98 events monitored, 43 met our quality-control criteria and were
 processed$^{24}$. Most of the efficiency comes from the 26 most
 densely covered light curves, which provide a representative and
 reliable sub-sample of events. We then computed the survey
 sensitivity for the whole time span 2002--07 by weighting each
 observing season relative to 2004, according to the number of events
 observed by PLANET for different ranges of peak magnification. This
 is described in the Supplementary Information, and illustrated in
 Supplementary Fig.~2. The resulting planet sensitivity is plotted in
 blue in Fig.~1, where the labelled contours show the corresponding
 expected number of detections. The figure shows that the core
 sensitivity covers 0.5--10~AU for masses between those of
 Uranus/Neptune and Jupiter, and extends (with limited sensitivity)
 down to about 5~$\Mearth$. As inherent to the microlensing technique,
 our sample of event-host stars probes the natural mass distribution
 of stars in the Milky Way (K--M dwarfs), in the typical mass range of
 0.14--1.0~$\Msun$ (see Supplementary Fig.~3).

 % ----------------------------------------------------------
 % FIGURE 2
 % ----------------------------------------------------------
 \begin{figure}[!h]
   \includegraphics[width=\columnwidth]{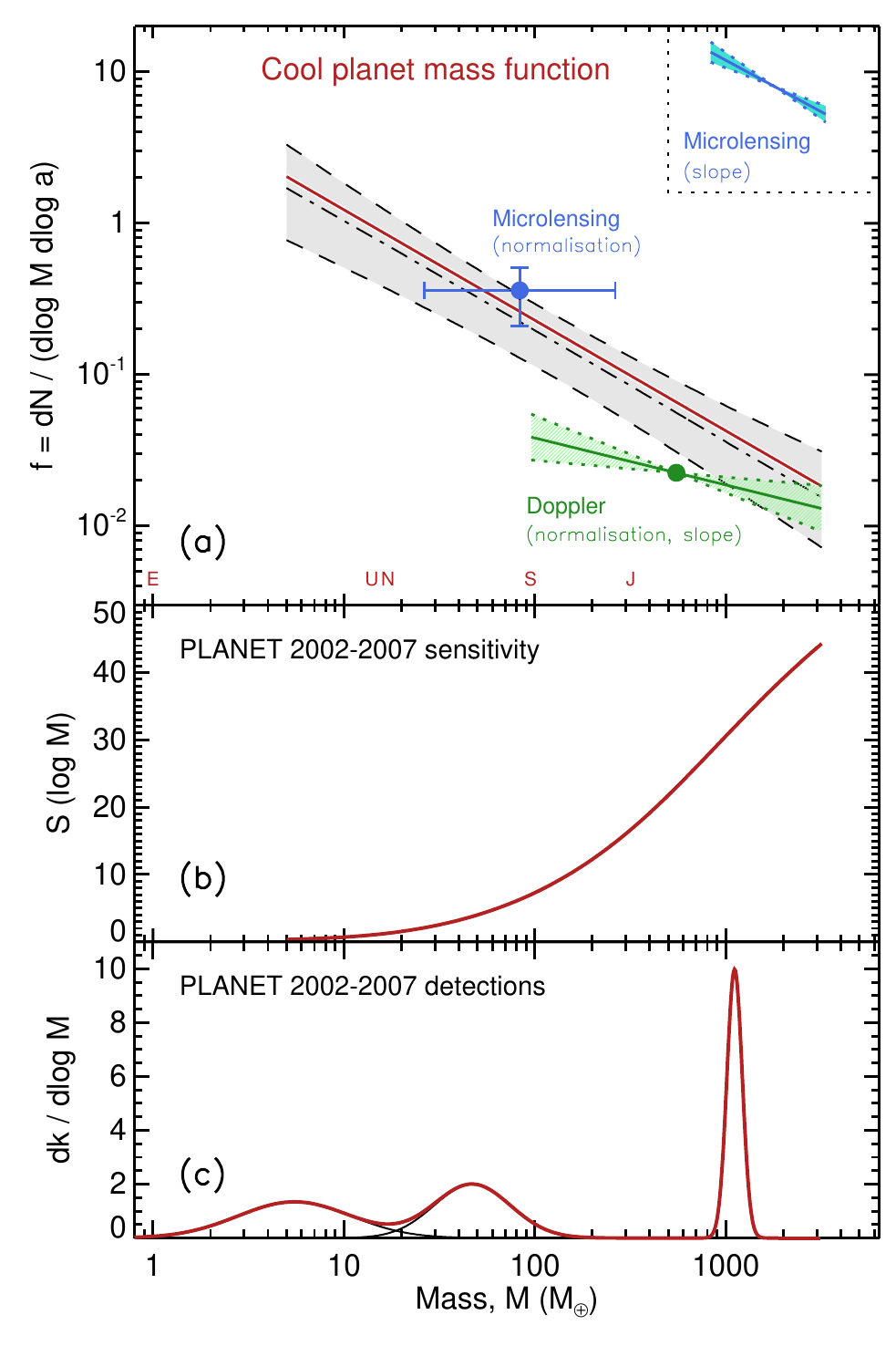}
   \caption{\textbf{Cool-planet mass function.} \textbf{a,} The
     cool-planet mass function, $f$, for the orbital range 0.5--10~AU
     as derived by microlensing. Red solid line, best fit for this
     study, based on combining the results from PLANET 2002--07 and
     previous microlensing estimates$^{18,25}$ for slope (blue line;
     error, light-blue shaded area, s.d.) and normalization (blue
     point; error bars, s.d.). We find $\dd N/(\dd\log a\ \dd\log M) =
     10^{-0.62 \pm 0.22}(M/M_{\rm Sat})^{-0.73 \pm 0.17}$, where $N$
     is the average number of planets per star, $a$ the semi-major
     axis and $M$ the planet mass. The pivot point of the power-law
     mass function is at the mass of Saturn ($M_{\rm Sat} = 95
     \,\Mearth$). Grey shaded area, 68\% confidence interval around
     the median (dash-dotted black line). For comparison, the
     constraint from Doppler measurements$^{27}$ (green line and
     point; error, green shaded area, s.d.) is also
     displayed. Differences can arise because the Doppler technique
     focuses mostly on solar-like stars, whereas microlensing a priori
     probes all types of host stars. Moreover, microlensing planets
     are located further away from their stars and are cooler than
     Doppler planets. These two populations of planets may then follow
     a rather different mass function. \textbf{b,} PLANET 2002--07
     sensitivity, $S$: the expected number of detections if all stars
     had exactly one planet, regardless of its orbit. \textbf{c,}
     PLANET 2002--07 detections, $k$. Thin black curves, distribution
     probabilities of the mass for the three detections contained in
     the PLANET sample; red line, the sum of these distributions.}
   \label{fig:cassan_fig2.pdf}
 \end{figure}
 % ----------------------------------------------------------

 To derive the actual abundance of exoplanets from our survey, we
 proceeded as follows. Let the planetary mass function, $f(\log a,
 \log M) \equiv \dd N/(\dd\log a\ \dd\log M)$, where $N$ is the
 average number of planets per star. We then integrate the product
 $f(\log a, \log M)\ S(\log a,\log M)$ over $\log a$ and $\log M$. This
 gives $E(f)$, the number of detections we can expect from our
 survey. For $k$ (fractional) detections, the model then predicts a
 Poisson probability distribution $P(k|E) = e^{-E}E^k/k!$. A Bayesian
 analysis assuming an uninformative uniform prior $P(\log f) \equiv 1$
 finally yields the probability distribution $P(\log f|k)$ that is
 used to constrain the planetary mass function.

 Although our derived planet-detection sensitivity extends overalmost
 three orders of magnitude of planet masses (roughly 5~$\Mearth$ to
 10~$\MJ$), it covers fewer than 1.5 orders of magnitude in orbit sizes
 (0.5--10~AU), thus providing little information about the dependence
 of $f$ on $a$. Within these limits, however, we find that the mass
 function is approximately consistent with a flat distribution in
 $\log a$ (that is, $f$ does not explicitly depend on $a$). The
 planet-detection sensitivity integrated over $\log a$, or $S(\log
 M)$, is displayed in Fig.~2b. The distribution probabilities of the
 mass for the three detections (computed according to the mass-error
 bars reported in the literature) are plotted in Fig.~2c (black
 curves), as is their sum (red curve).

 To study the dependence of $f$ on mass, we assume that to the first
 order, $f$ is well-approximated by a power-law model: $f_0\
 (M/M_0)^\alpha$, where $f_0$ (the normalization factor) and $\alpha$
 (the slope of the power-law) are the parameters to be derived and
 $M_0$ a fiducial mass (in practice, the pivot point of the mass
 function). Previous works$^{18,25-27}$ on planet frequency have
 demonstrated that a power law provides a fair description of the
 global behaviour of $f$ with planetary mass. Apart from the
 constraint based on our PLANET data, we also made use in our analysis
 of the previous constraints obtained by microlensing: an estimate of
 the normalization$^{18}$ $f_0$ ($0.36 \pm 0.15$) and an estimate of
 the slope$^{25}$ ($-0.68 \pm 0.2$), displayed respectively as the
 blue point and the blue lines in Fig.~2. The new constraint presented
 here therefore relies on 10 planet detections. We obtained $10^{-0.62
   \pm 0.22}\ (M/M_0)^{-0.73 \pm 0.17}$ (red line in Fig.~2a) with a
 pivot point at $M_0 \simeq 95\ \Mearth$; that is, at Saturn's mass.
 The median of $f$ and the 68\% confidence interval around the median
 are marked by the dashed lines and the grey area.  

 Hence, microlensing delivers a determination of the full planetary
 mass function of cool planets in the separation range 0.5--10~AU. Our
 measurements confirm that low-mass planets are very common, and that
 the number of planets increases with decreasing planet mass, in
 agreement with the predictions of the core-accretion theory of planet
 formation$^{28}$. The first microlensing study of the abundances of
 cool gas giants$^{21}$ found that fewer than 33\% of M dwarfs have a
 Jupiter-like planet between 1.5--4 AU, and even lower limits of 18\%
 have been reported$^{29,30}$. These limits are compatible with our
 measurement of $5_{-2}^{+2}\%$ for masses ranging from Saturn to 10
 times Jupiter, in the same orbit range.
 
 From our derived planetary mass function, we estimate that within
 0.5--10~AU (that is, for a wider range of orbital separations than
 previous studies), on average $17_{-9}^{+6}\%$ of stars host a
 'Jupiter' (0.3--10~$\MJ$), and $52_{-29}^{+22}\%$ of stars host
 Neptune-like planets (10--30~$\Mearth$). Taking the full range of
 planets that our survey can detect (0.5--10~AU, 5~$\Mearth$ to
 10~$\MJ$), we find that on average every star has
 $1.6_{-0.89}^{+0.72}$ planets. This result is consistent with every
 star of the Milky Way hosting (on average) one planet or more in an
 orbital-distance range of 0.5--10~AU. Planets around stars in our
 Galaxy thus seem to be the rule rather than the exception.
 
 % -----------------------------
 % Letter - References
 % -----------------------------
 
 \nocite{MayorQueloz95}  %   1
 \nocite{MarcyButtler1996} %  2
 \nocite{Charbonneau2000} %  3
 \nocite{Howard2011} %  4
 \nocite{Mayor2011} %  5
 \nocite{MaoPaczynski1991} %  6
 \nocite{GouldLoeb1992} %  7
 \nocite{BennettRhie1996} %  8
 \nocite{Wambsganss1997} %  9
 \nocite{SumiUnbound2011} %  10
 \nocite{Ref-OGLE} %  11
 \nocite{Ref-MOA} %  12
 \nocite{PilotPLANET} %  13
 \nocite{OB05169Lb}  %  14
 \nocite{OGLE07109Lbc} %  15
 \nocite{OB05071Lb}  %  16
 \nocite{Dong2008ob71} %  17
 \nocite{Gould4years2010} %  18
 \nocite{OGLE05390Lb} %  19
 \nocite{Jovi2008} %  20
 \nocite{Gaudi5years} %  21
 \nocite{Einstein1936} %  22
 \nocite{DominikGalactik} %  23
 \nocite{Causfix} %  24
 \nocite{SumiOGLE368} %  25
 \nocite{Howard2010} %  26
 \nocite{Cumming2008PASP} %  27
 \nocite{Pollack1996} %  28
 \nocite{Tsapras2003} %  29
 \nocite{Snodgrass2004} %  30
 
 % -----------------------------
 % References
 % -----------------------------
 
 {\small
   \bibliographystyle{moainature}
   \bibliography{newmoai}
 }
 
 % -----------------------------
 % Informations
 % -----------------------------

 \begin{small}
   
   \paragraph{Supplementary Information}
   is linked to the online version of the paper at
   www.nature.com/nature.

   \paragraph{Acknowledgements}
   Support for the PLANET project was provided by the French National
   Centre for Scientific Research (CNRS), NASA, the US National
   Science Foundation, the Lawrence Livermore National
   Laboratory/National Nuclear Security Administration/Department of
   Energy, the French National Programme of Planetology, the Program
   of International Cooperation in Science France--Australia,
   D. Warren, the German Research Foundation, the Instrument Center
   for Danish Astronomy and the Danish Natural Science Research
   Council. The OGLE collaboration is grateful for funding from the
   European Research Council Advanced Grants
   Program. K.Ho. acknowledges support from the Qatar National
   Research Fund. M.D. is a Royal Society University Research Fellow.

   \paragraph{Author Contributions}
   A.Ca. led the analysis and conducted the modelling and statistical
   analyses. A.Ca. and D.K. selected light curves from 2002--07
   PLANET/OGLE microlensing seasons, analysed the data and wrote the
   Letter and Supplement. D.K. computed the magnification maps used
   for the detection-efficiency calculations. J.-P.B. and Ch.C. wrote
   the software for online data reduction at the
   telescopes. J.-P.B. led the PLANET collaboration, with M.D., J.G.,
   J.M. and A.W.; P.F. and M.D.A. contributed to online and offline
   data reduction. M.D. contributed to the conversion of the detection
   efficiencies to physical parameter space and developed the PLANET
   real-time display system with A.W., M.D.A. and Ch.C.; K.Ho. and
   A.Ca. developed and tested the Bayesian formulation for fitting the
   two-parameter power-law mass function. J.G. edited the manuscript,
   conducted the main data cleaning and managed telescope operations
   at Mount Canopus (1~m) in Hobart. J.W. wrote the original
   magnification maps software, discussed the main implications and
   edited the manuscript. J.M., A.W. and U.G.J. respectively managed
   telescope operations in South Africa (South African Astronomical
   Observatory 1~m), Australia (Perth 0.61~m) and La Silla (Danish
   1.54~m). A.U. led the OGLE campaign and provided the final OGLE
   photometry. D.P.B, V.B., S.B., J.A.R.C., A.Co., K.H.C., S.D.,
   D.D.P., J.D., P.F., K.Hi., N.K., S.K., J.-B.M., R.M., K.R.P.,
   K.C.S., C.V., D.W., B.W. and M.Z. were involved in the PLANET
   observing strategy and/or PLANET data acquisition, reduction,
   real-time analysis and/or commented on the
   manuscript. T.S. commented on the manuscript. M.K.S., M.K., R.P.,
   I.S., K.U., G.P. and L.W. contributed to OGLE data.

   \paragraph{Author Information}
   Reprints and permissions information is available at
   www.nature.com/reprints. The authors declare no competing financial
   interests. Readers are welcome to comment on the online version of
   this article at www.nature.com/nature. Correspondence and requests
   for materials should be addressed to A.Ca. (cassan@iap.fr).
   
 \end{small}

\end{document}